\documentstyle[psfig,conf_iap,10pt]{article}
\def\jccplotmacro#1#2#3#4#5#6#7{\centering \leavevmode
    \vbox to#2{\rule{0pt}{#2}}
    \includegraphics{#1}}
\begin{document}
\heading{%
%
The Distribution of Gas In and Around Galaxies
%
} 
\par\medskip\noindent
\author{%
Jane C. Charlton and Christopher W. Churchill
}
\address{%
Astronomy and Astrophysics Department
525 Davey Laboratory
The Pennsylvania State University
University Park, PA 16802
}

\begin{abstract}
A consensus is developing on
the nature of various populations of absorbers at
different redshift regimes and in particular
on their relationships to galaxies at those epochs.
As one example we discuss the population of $z<1$
Mg~{\sc ii} absorbers.  Kinematic models are presented that
show this population to be consistent with some
combination of halo and disk kinematics and not with
a pure disk or halo model.  In contrast, the $2 < z < 3$
C~{\sc iv} absorbers are likely to represent a mix of already
formed halos and protogalactic clumps in the process
of assembly.  Clues could be provided by the level at
which S~{\sc iv} traces the same kinematic components
as C~{\sc iv} (relative to how it traces lower ionization gas)
and this is illustrated by a simple application of
principal components analysis.
It remains uncertain how much unambiguous information
it will be possible to extract about the detailed
properties of the gas along the line of sight through
an individual absorption profile.  However, 
ratios of various low and high ionization transitions
along the line of sight can be
combined with insights gained from kinematic models and
from observations of galaxy properties.
There is reason for optimism that we will progress toward
a point when we will someday study the interstellar medium and
gaseous halos and environments of absorbing galaxies
at close to the same level as we study these components
in the Milky Way.

\end{abstract}
\section{Introduction}
The past decade of research in quasar absorption lines
had the theme of establishing the basic nature of
various classes of absorbers, such as the Ly$\alpha$
forest clouds, the damped Ly$\alpha$ absorbers, the
C~{\sc iv} systems, and the Mg~{\sc ii} systems.  The emphasis
has been on the relationship between a population of absorbers
and the population of galaxies.  For example, common
questions are whether they are produced by a particular 
component of galaxies, and whether absorbers are
clustered with respect to galaxies?  The answer
for a particular population can vary as a function 
of time and thus quasar absorption lines provide a
basic understanding of the assembly of gas into
galaxies and of the evolution of the gaseous content
of the Universe.
Although significant controversy remains (eg. the 
high redshift damped Ly$\alpha$ clouds, see both 
Wolfe and Haehnelt in this volume), we can
be confident that this basic understanding will
soon be achieved.  

Even with this significant advance, a major potential
of quasar absorption lines remains untapped.
Absorption line studies of the Milky Way Galaxy
provide detailed information about its gaseous components,
eg. disk vs. halo composition, high velocity clouds, and
supershells.  Will quasar absorption line studies ever
allow such detailed information to be extracted about
other individual galaxies?  In any one case, can we
understand the physical conditions in internal 
structures present along a line of sight through a
galaxy?


\begin{figure}[t]
\jccplotmacro{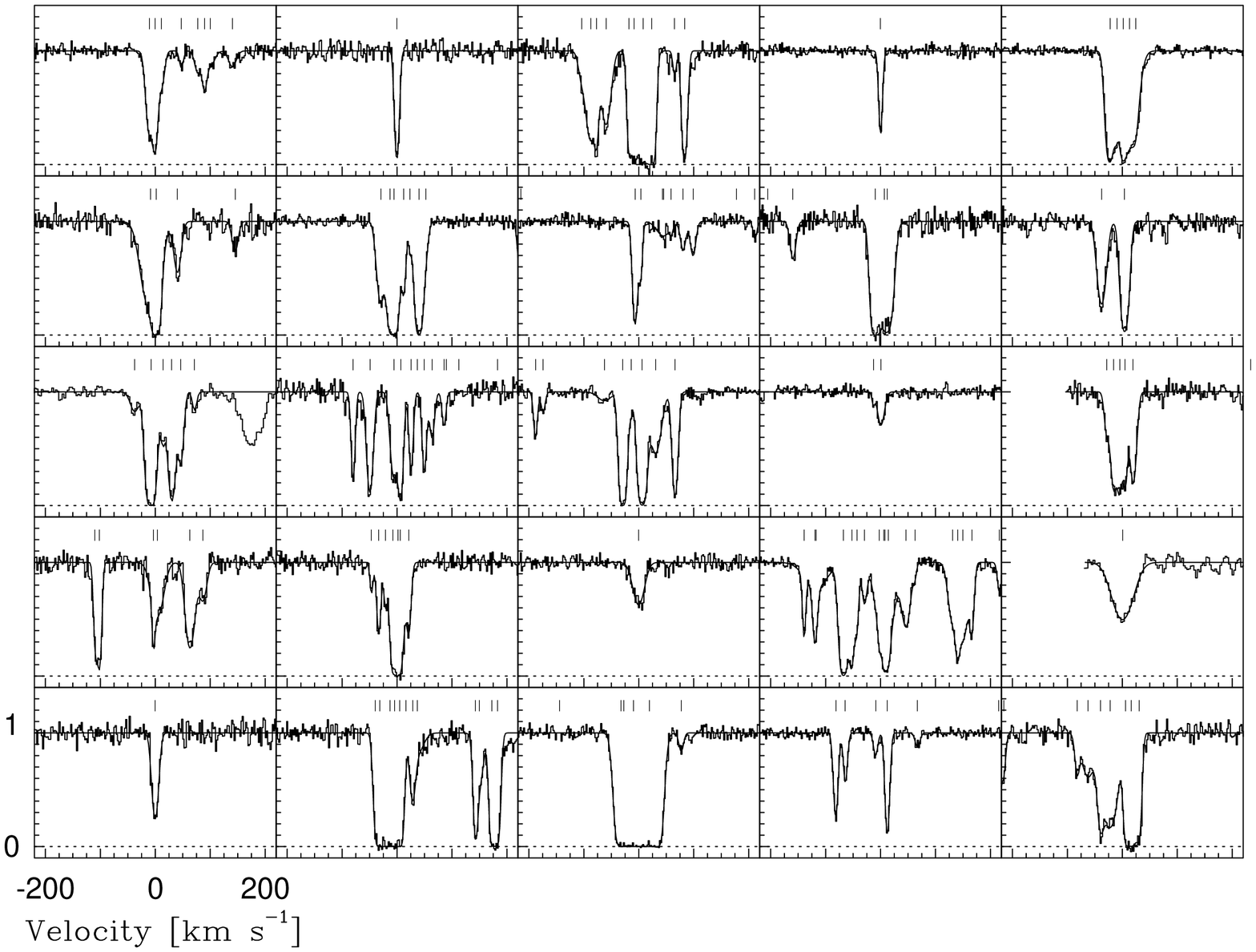}{3.5in}{0}{50.}{40.}{-214}{49}
\vglue -1.2in
\caption{An ensemble of Mg~{\sc ii}($\lambda$ 2796) absorption profiles
for absorbers at $0.4 < z < 1.0$ at resolution 6~km~s$^{-1}$ from 
Keck/HIRES.  The Voigt profile model fits are superimposed
upon the data and the locations of those components are 
marked with ticks.
}
\end{figure}

\section{Diagnosing the Conditions in Mg~{\sc ii} Absorbers at $z<1$}

As our first example we present in Fig.~1 Keck/HIRES spectra
of the Mg~{\sc ii}($\lambda$ 2796) profiles for a population of Mg~{\sc ii} 
absorbers at $0.4 < z < 1.0$.  We would like to be able
to identify in these profiles the signatures of spirals
and ellipticals, of disk/halo interplay, of satellite
galaxies, of supershells, etc.  This section discusses
the ensemble of absorbing galaxies that produce
the variety of profiles displayed
and sketches out some considerations that have
potential for diagnosing their individual properties.

We have generated ensembles of simulated spectra for lines
of sight through model populations of galaxies with various
spatial and kinematic distributions of Mg~{\sc ii} gas.
The details of these models and the results will be
presented in Charlton and Churchill (1997).
In each case a galaxy was created by placing some number of 
clouds in a halo and/or disk component, with the velocity vector
of each cloud chosen from a given distribution, eg. rotation
or radial infall with an additional random component.
The choice of which populations of galaxies were
considered in our models
was guided by what we know about the properties of nearby
galaxies.  The cloud Mg~{\sc ii} column densities and Doppler
$b$ parameters were chosen from input distribution functions
so as to produce output cloud properties that match
those extracted from the ensemble of observed $z<1$ Mg~{\sc ii}
profiles presented in Fig.~1.  In this way we
remove the ambiguities caused by lack of 
knowledge of chemical abundance and ionization
conditions and are able to focus on distinguishing
spatial and kinematic distributions of clouds.

\begin{figure}[thb]
\jccplotmacro{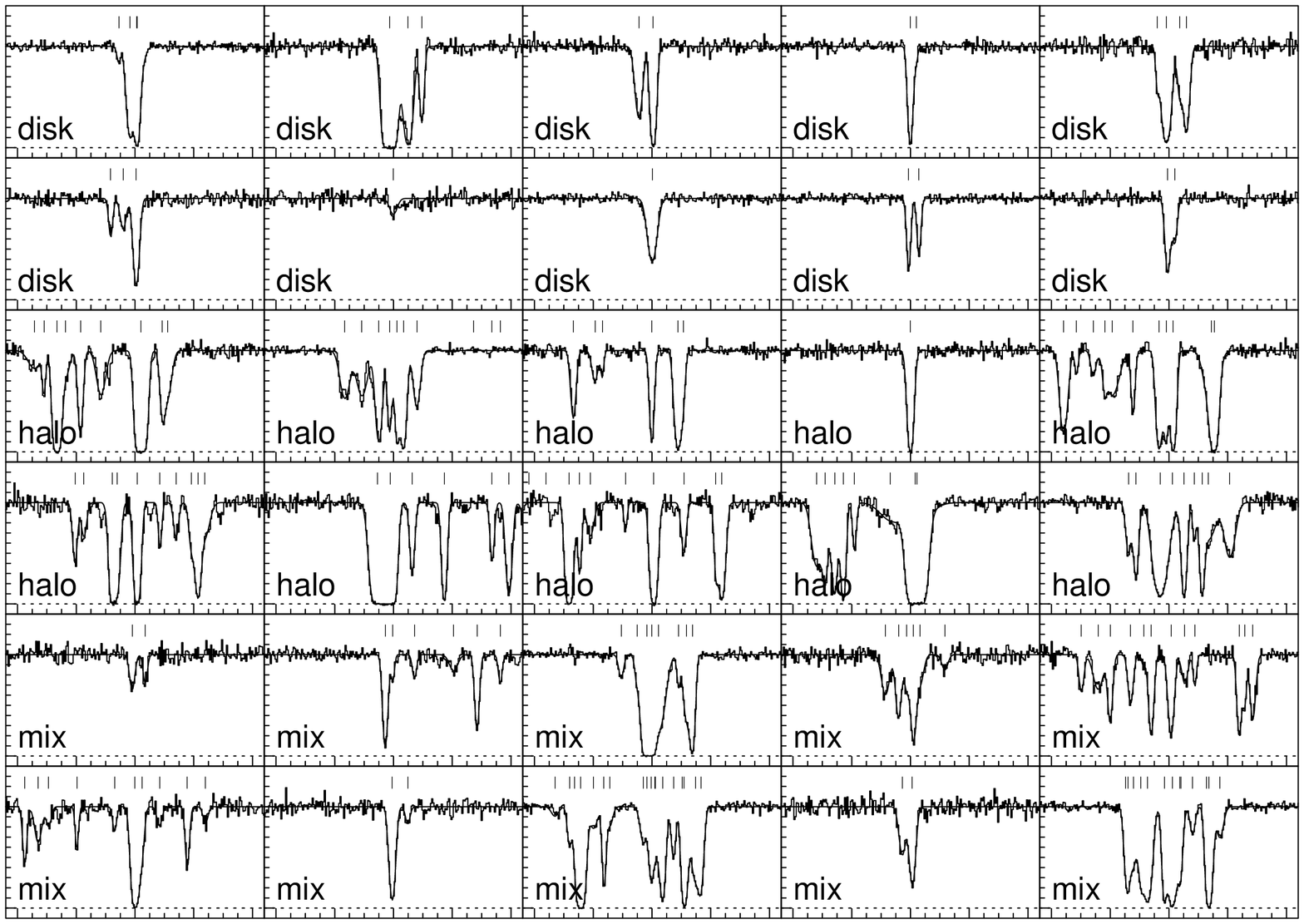}{3.5in}{0}{50.}{40.}{-214}{49}
\vglue -1.3in
\caption{Ensembles of simulated spectra for three
different kinematic models.  The parameters of these
models were tuned in order to reproduce the flux distribution
of the observed $z<1$ Mg~{\sc ii} profiles.  For the first ten
profiles, all clouds were placed
within a thick rotating disk that becomes thicker with
increasing radius, and that has a velocity dispersion
of $30$~km~s$^{-1}$ in the vertical direction.  For the second
set of ten, all clouds are spread at random in a spherical
halo and have constant magnitude radial infall, and for
the ``mix'' model half the clouds were placed in the disk and 
half in the halo.
}
\end{figure}

Fig.~2 presents examples of profiles
generated from lines of sight through rotating disks 
and through halos with radial infall.  In most cases we
could guess from which kinematic law a given profile was 
chosen, but in some cases we cannot.  It is clear
from inspection of these profiles that neither of
these simple pure disk or pure halo models produces
sufficient variety to match the observed ensemble.
The pure halo models do not produce a large enough
fraction of simple, narrow ($< 100$~km~s$^{-1}$) profiles,
and the pure disk models do not produce enough complex
profiles with large velocity spreads.  An ensemble
of model galaxies in which half the Mg~{\sc ii} clouds are in 
the disk and half are in the halo produces a much
more realistic set of profiles.  We performed many
statistical tests to distinguish what types of
kinematic models best fit the data.  For example,
the two point clustering function is defined as the
distribution of differences in velocity between all 
pairs of subcomponents (the ticks in Fig.~1 and 2).
This function is too narrow compared to the data for
pure disk models and too wide for pure halo models.
This test and others
showed that some Mg~{\sc ii} clouds must exist in
a narrow kinematic distribution such as a disk and
others in a wider distribution in order to match
the observations.  It is interesting that
agreement can be obtained either from a model in 
which half of the clouds in each galaxy are in the
halo and half in the disk or from a model in which
half the galaxies had pure disk and half had
pure halo contributions.  In reality the galaxy
population responsible for Mg~{\sc ii} absorbers is
likely to be a continuum of relative halo/disk
proportions just as would be expected from
the gas distributions in nearby galaxies.

%
\begin{figure}[thb]
\jccplotmacro{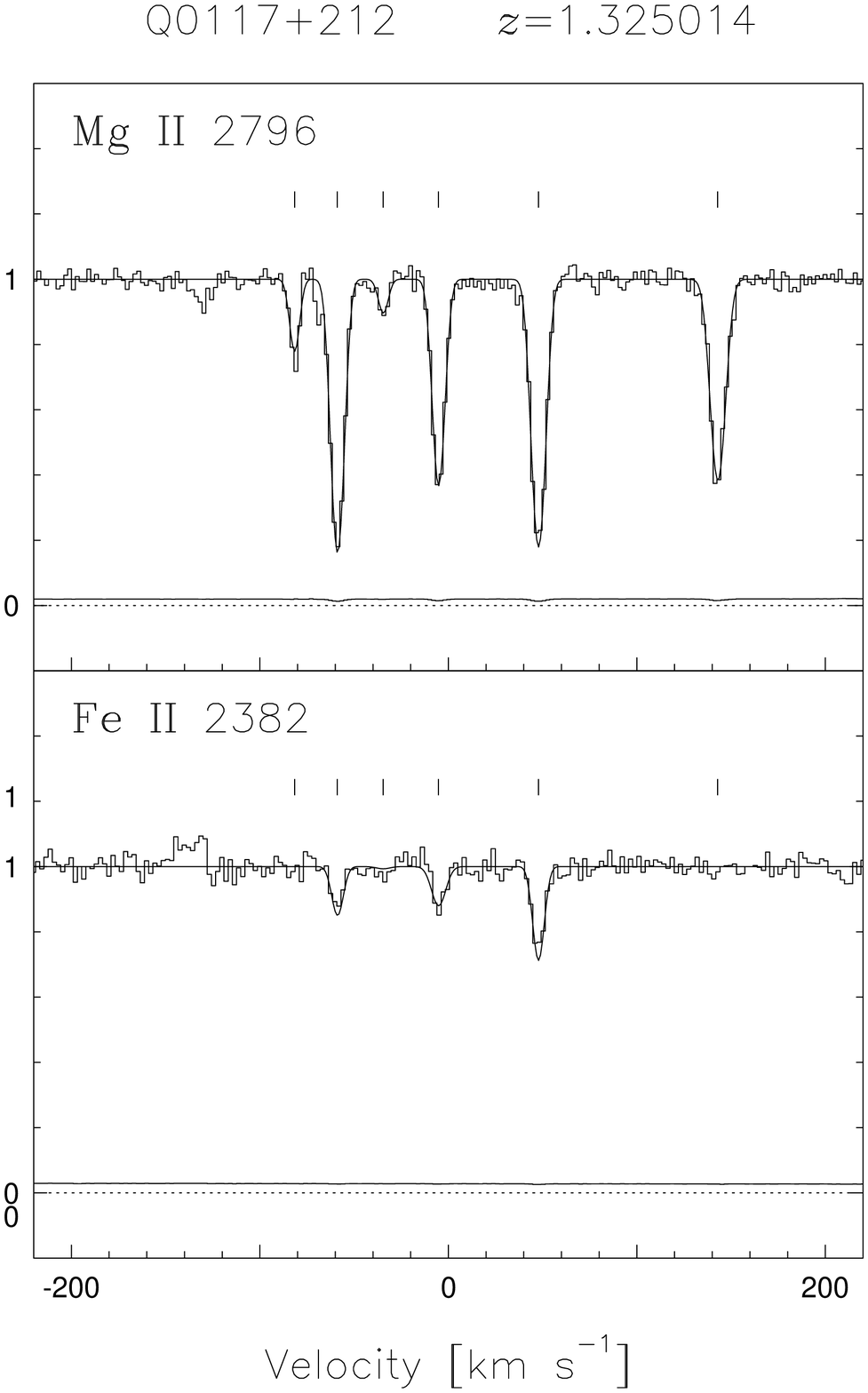}{1.85in}{0}{25.}{25.}{-153}{-43}
\jccplotmacro{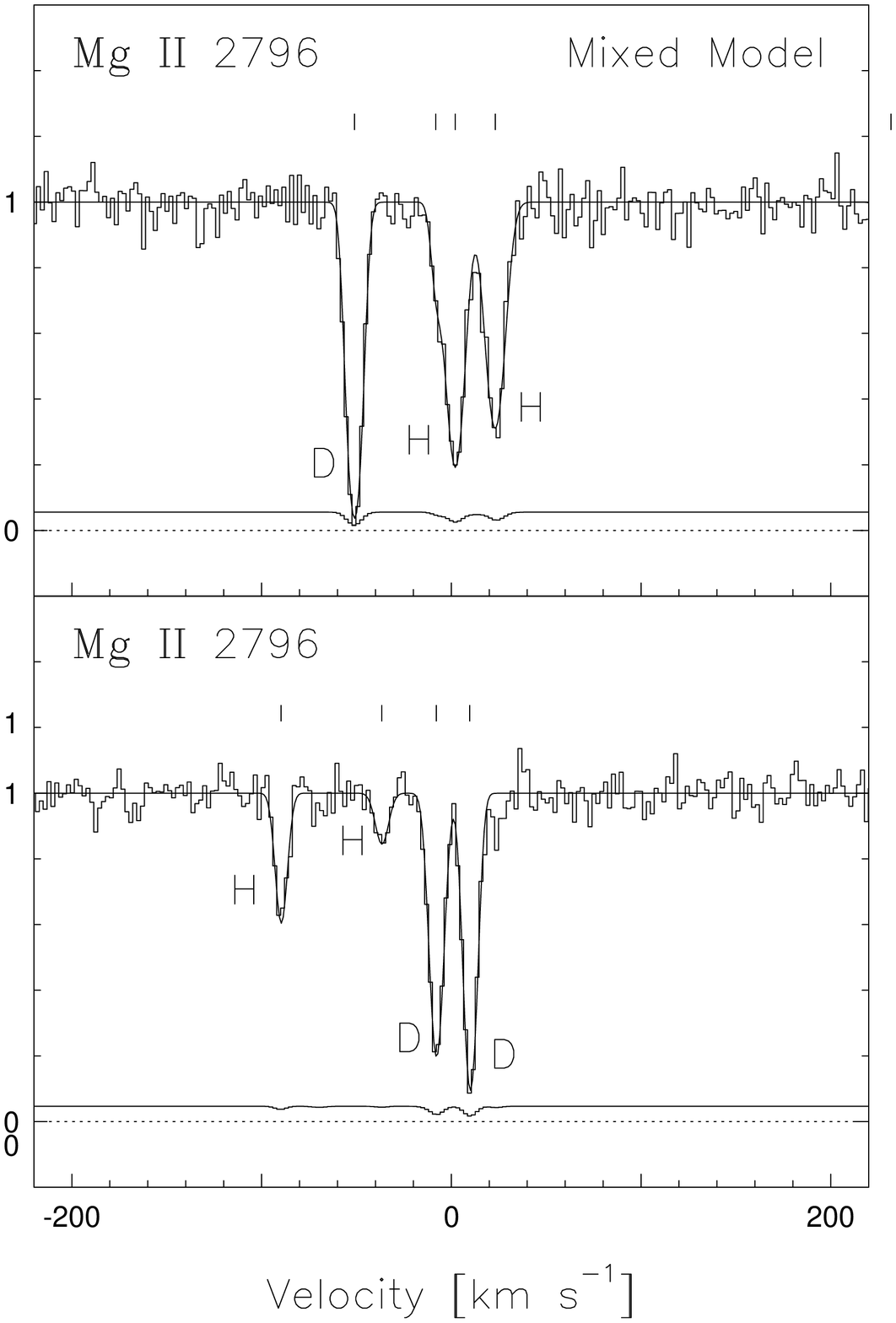}{1.85in}{0}{25.}{25.}{0}{103}
\vglue -1.72in
\caption{On the left, the Keck/HIRES absorption profile showing the
variation of Fe~{\sc ii}/Mg~{\sc ii} with velocity in the
$z=1.3250$ system of Q0117+212.  (Note the small 
Fe~{\sc ii}/Mg~{\sc ii} at $+140$~km~s$^{-1}$.)  On the right, two simulated
spectra of Mg~{\sc ii}($\lambda$ 2796) profiles from a combined halo/disk
model with the place of origin identified for individual clouds.
This shows that without additional information there is
considerable ambiguity in the kinematic profiles.
}
\end{figure}

Monte--Carlo variations between different lines of
sight through galaxies drawn from the same model
parameters can go a long way toward explaining
the variations in observed profiles of the $z<1$
Mg~{\sc ii} population.  The variations
between different galaxies may not be so large
as the variations within them.  Thus it appears
that in any one case we cannot definitively extract 
the nature of the absorbing galaxy
from the profiles of low ionization gas.
However, additional information provided
by the ratio of Fe~{\sc ii}/Mg~{\sc ii} has the potential
to provide more leverage.  The example in Fig.~3
shows that the ratio can vary dramatically across
individual systems.  The observed ratios may allow
us to distinguish halo clouds
from disk clouds, and starbursting dwarfs (smaller
Fe~{\sc ii}/Mg~{\sc ii}) or quiescent dwarfs 
(larger Fe~{\sc ii}/Mg~{\sc ii})
from material associated with a bright galaxy.
Information from deep images of the
field can provide considerable further leverage.
Finally, Churchill has shown (in these proceedings)
that the distribution of high ionization gas can
provide information about the kinematic components
of $z<1$ Mg~{\sc ii} absorbers.  The velocity spread of
the Mg~{\sc ii} profiles tends to be larger (at a 3.8$\sigma$
level) for systems with a larger C~{\sc iv} equivalent
width, suggesting a picture of an extended C~{\sc iv}
halo with embedded Mg~{\sc ii} clouds.  The additional
fact that the Mg~{\sc ii} equivalent width is not so
strongly correlated with the Mg~{\sc ii} kinematics
leads us to consider that a dominant fraction of 
the Mg~{\sc ii} column may be produced in the galaxy disk.
HST/STIS observations at high resolution
are needed to determine just how much of the 
C~{\sc iv} is in the disk, in the halo, and in outer 
layers surrounding individual Mg~{\sc ii} clouds.

\section{Diagnosing the Conditions in C~{\sc iv} Absorbers at $z>2$}

As our second example, we briefly consider the population of C~{\sc iv}
absorbers at $z>2$.  For these systems the issue is whether
a given one is a collection of protogalactic clumps expanding
with the Hubble flow, or part of an already collapsed structure
\cite{RHS}.  As with the Mg~{\sc ii} absorbers, an important question
is whether one can distinguish these two possibilities by
examining an individual set of profiles for a C~{\sc iv} system.
Here the most important clue could be whether the
Si~{\sc iv} (an intermediate ionization state transition)
is distributed more like the lower ionization transitions
(such as Mg~{\sc ii}, C~{\sc ii}, or Fe~{\sc ii}) or like the higher ionization
transitions (such as C~{\sc iv} or N~{\sc v}).  Here we present a
technique for quantifying the degree to which two ions
trace one another in velocity space.  Principal components
analysis allows us to represent the observed transitions
in terms of a set of basis functions \cite{Fra}.  A successful PCA
analysis will describe the variations between the transitions
with many fewer basis functions than there are transitions.
This technique is illustrated in Fig.~4
for two example C~{\sc iv} systems.  Most of the variation 
between the profiles can
be explained by two principal components that resemble
the profiles for the C~{\sc iv} and for a lower ionization
species, respectively.  In one system the Si~{\sc iv} is
distributed similarly to the C~{\sc ii}, and this can be
quantified by considering the coefficients of the
principal components.  In the second system, the
Si~{\sc iv} has nearly identical principal component
coefficients as C~{\sc iv}.  As with the Fe~{\sc ii}/Mg~{\sc ii} 
ratios, the Si~{\sc iv}/C~{\sc iv} ratio is found to vary across some
profiles.  This can be diagnostic of changing abundance
ratios, ionization parameters, or spectral shapes (with
large values for stellar rather than power law), however
again there is considerable ambiguity (see also Boksenberg
in this volume).

%
\begin{figure}[thb]
\jccplotmacro{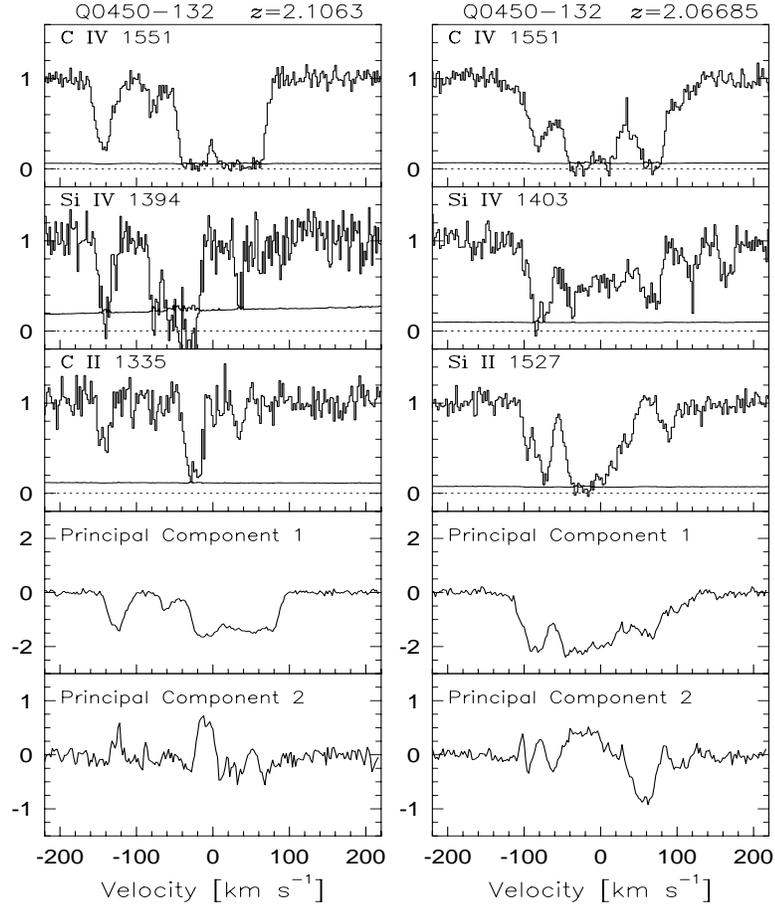}{5.5in}{0}{60.}{50.}{-200}{70}
\vglue -1.44in
\caption[]{Two examples of principal components analysis.
Only some of the transitions used are displayed.}
\end{figure}

\section{Future Plans for Investigating $z>2.5$ Mg~{\sc ii} Absorbers}

The properties of $z<1$ Mg~{\sc ii} absorbers are consistent with
clouds within the usual kinematic components of galaxies,
i.e. halos and disks.  C~{\sc iv} absorbers at high redshift show
signatures of coalescence of protogalactic clumps.
Do the higher redshift Mg~{\sc ii} absorbers already show
kinematic signatures of disks, or are they more
consistent with the distribution of higher ionization
gas?  This question can only be addressed with near--IR
spectroscopy for the $z>2.5$ Mg~{\sc ii} absorbers.  Using
the JCAM, a $R=10,000$ spectrograph planned for the 
8--m Hobby--Eberly Telescope, we will conduct a survey
for Mg~{\sc ii} doublets to a rest equivalent width limit of
0.15~\AA.  For detected systems we plan to determine
the kinematic structure of Mg~{\sc ii} systems up to a $z=4$
using a $R=20,000$ mode of the JCAM.  In this important
redshift regime we may find evidence for the assembly
of galaxies and for the gradual development of their
gaseous components.

\acknowledgements{The National Science Foundation has supported this
work through AST--9617185.  Thanks to Paul Francis for his PCA code
and to Rajib Ganguly for his contributions to the C~{\sc iv} analysis.}


%
\vfill

\begin{iapbib}{99}{
\bibitem{Cha} Charlton, J. C., \& Churchill, C. W., 1997, \apj, submitted
\bibitem{Fra} Francis, P. J., Hewett, P. C., Foltz, C. B., \& Chaffee, F. H.,
              1992, \apj, 398, 476
\bibitem{RHS} Rauch, M., Haehnelt, M. G., \& Steinmetz, M.,
              1997, \apj, 481, 601
%
}
\end{iapbib}
\end{document}